\documentclass[journal=jctcce,manuscript=article,layout=traditional]{achemso}
\usepackage{fullpage}
\usepackage{mathtools}
\usepackage{amsthm}
\usepackage{amssymb}
\usepackage{bm}
\usepackage[version=3]{mhchem}
\usepackage{braket}
\usepackage{relsize}
\usepackage{enumitem}
\usepackage[font=small,labelfont=bf]{caption}
\usepackage{graphicx}
\usepackage{footnote}
\usepackage{tabularx}
\usepackage{booktabs}
\usepackage{makecell}
\usepackage{threeparttable}
\usepackage{chemfig}

\usepackage{longtable}
\usepackage[mathscr]{euscript}
\usepackage{threeparttable}  

\usepackage{natbib}
\usepackage{multicol}

\usepackage{relsize}
\usepackage{tabularx}
\usepackage{tablefootnote}

\newcommand{\RNum}[1]{\expandafter{\romannumeral #1\relax}}

\newcolumntype{C}{>{\centering\arraybackslash}X}

\title{Gold-Standard Chemical Database 137 (GSCDB137): A diverse set of accurate energy differences for assessing and developing density functionals}	

\author{Jiashu Liang}
\affiliation{
	Kenneth S. Pitzer Center for Theoretical Chemistry,
	Department of Chemistry,
	University of California at Berkeley,
	Berkeley, CA 94720, USA
}
\author{Martin Head-Gordon}
\affiliation{
	Kenneth S. Pitzer Center for Theoretical Chemistry,
	Department of Chemistry,
	University of California at Berkeley,
	Berkeley, CA 94720, USA
}
\alsoaffiliation{
	Chemical Sciences Division,
	Lawrence Berkeley National Laboratory,
	Berkeley, CA 94720, USA
}
\email{mhg@cchem.berkeley.edu}

\date{\today}

\begin{document}

\begin{tocentry}
\includegraphics[width=8cm, height=4.3 cm]{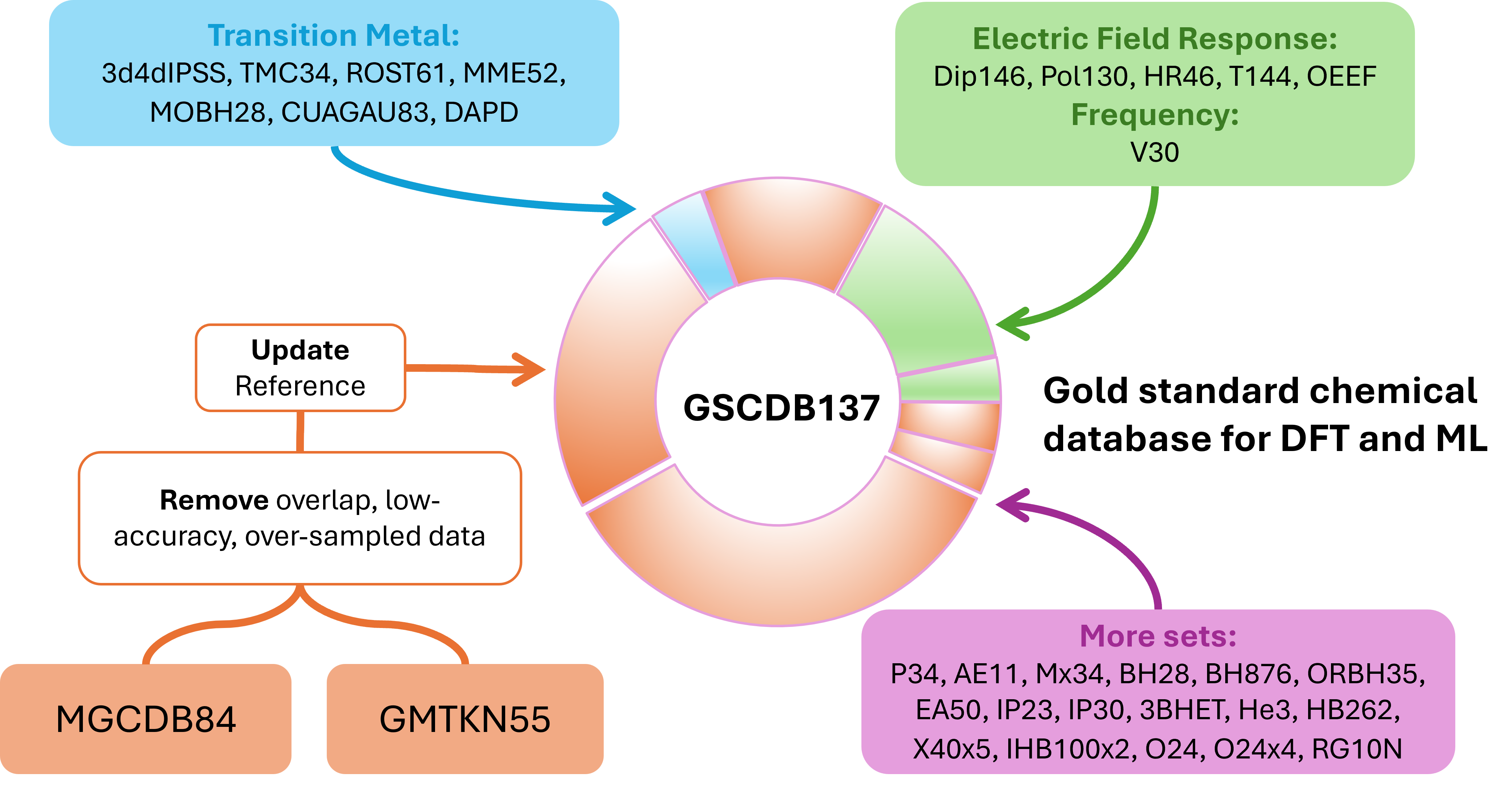}
\end{tocentry}

\begin{abstract}
We present GSCDB137, a rigorously curated benchmark library of 137 data sets (8377 entries) covering main-group and transition-metal reaction energies and barrier heights, (intramolecular) non-covalent interactions, dipole moments, polarizabilities, electric-field response energies, and vibrational frequencies. Legacy data from GMTKN55 and MGCDB84 have been updated to today’s best reference values; redundant, spin-contaminated, or low-quality points were removed, and many new, property-focused sets were added. Testing 29 popular density functional approximations (DFAs) confirms the expected Jacob’s-ladder hierarchy overall but also reveals notable exceptions: functional performance for frequencies and electric-field properties correlates poorly with that for other ground-state energetics. $\omega$B97M-V and $\omega$B97X-V are the most balanced hybrid meta-GGA and hybrid GGA, respectively; B97M-V and revPBE-D4 lead the meta-GGA and GGA classes. Double hybrids lower mean errors by about 25 \% versus the best hybrids but demand careful frozen-core, basis-set, and multi-reference treatment. GSCDB137 offers a comprehensive, openly documented platform for stringent DFA validation and for training the next generation of non-empirical and machine-learned functionals.
\end{abstract}

\maketitle

\clearpage

\section{Introduction}
    
Exact electronic structure theory formally scales exponentially with the number of electrons and polynomially with a high order in the size of the atomic orbital (AO) basis set. Therefore, despite considerable progress~\cite{sherrill1999configuration,szalay2012multiconfiguration,tubman2020modern,eriksen2020shape}, exact methods at the complete basis set (CBS) limit remain computationally infeasible for realistic molecular systems. Coupled cluster (CC) theory,~\cite{bartlett2007coupled,bartlett2024perspective} however, provides a pragmatic high-accuracy alternative with more manageable scaling. With the judicious choice of excitation level truncation (perturbative triples or better) and careful treatment of the CBS limit via extrapolation~\cite{helgaker1997basis} or F12-type methods~\cite{knizia2009simplified}---can yield benchmark-level accuracy for molecular energy differences. Best practices for benchmarking have been discussed in several reviews~\cite{karton2017w4,Goerigk2017,rezac2020non,karton2025good}, and as we review briefly and necessarily incompletely below, there has been continuous progress in both accuracy and chemical diversity.

Given the prohibitive cost of high-level CC theory for large or complex systems, such as catalytic cycles or protein–ligand binding, an alternative is required for routine applications. Density functional theory (DFT)~\cite{kohn1965self,kohn1996density,capelle2006birdseyeviewdensityfunctionaltheory,mardirossian2017thirty} offers a more affordable option, scaling at most cubically with system size and achieving rapid convergence toward the CBS limit (exponential in the AO basis cardinal number)~\cite{dunning1989gaussian,feller1993use,helgaker1997basis}. However, there are many hundreds of approximate functionals available, and no single functional is universally reliable. Consequently, benchmark studies evaluating DFT performance using high-accuracy reference data—typically from CC theory—are essential to guide functional selection~\cite{Goerigk2017,mardirossian2017thirty}. Additionally, efforts to reduce the cost of DFT, such as composite methods~\cite{grimme2015consistent} or machine-learned surrogate models~\cite{behler2016perspective}, are trained using benchmark-quality data from either CC theory~\cite{smith2020ani} or from functionals validated against CC-level benchmarks~\cite{smith2017ani}..

Early benchmarks often used back-corrected experimental data to yield accurate electronic energies. The Gaussian-n (n=2-3) data sets~\cite{curtiss1997assessment,curtiss2000assessment,curtiss2005assessment}, consisting of properties such as atomization energies, ionization potentials, and electron affinities, were built this way in the early 1990s. These data sets are historically important because they were used to parameterize early groundbreaking density functionals such as B3LYP~\cite{Becke1993,stephens1994ab}. This line of work, in which experimental values are combined with computational methods to yield thermochemical data, continues today, particularly through the active thermochemical tables project~\cite{ruscic2004introduction}, which has managed to assign uncertainties to such data and pinpoint areas needing improvement.

As high-accuracy CC methods such as CCSD(T)~\cite{raghavachari1989fifth} became available in the early 1990s, and as computer performance continued to roughly double every two years, an explosion of activity followed to expand and broaden the categories of molecular energies with benchmark-level accuracy. Because chemical kinetics relies on reliable barrier heights, dedicated barrier height data sets for small molecules soon appeared~\cite{zhao2005multi,zhao2005benchmark,Zheng_2007}. Intermolecular interactions emerged as a second major focus: the seminal S22 data set~\cite{jurevcka2006benchmark}, later refined in both geometry and energetics~\cite{marshall2011basis}, inspired many subsequent non-covalent interaction benchmarks~\cite{Rezac2011_1,rezac2020non} and has been reviewed comprehensively~\cite{rezac2016benchmark}. These reference data sets enabled rigorous evaluations of density-functional performance and guided the design of improved functionals.

As more research groups began generating high-quality (and sometimes low-quality) benchmark data, the need to systematically integrate and curate these efforts into a larger, validated, and community-accessible database became increasingly apparent. A landmark effort was the GMTKN24 and GMTKN30 databases~\cite{goerigk2010general,goerigk2011efficient}, which cover general main-group thermochemistry, kinetics, and noncovalent interactions. In 2017, the larger GMTKN55 compilation was introduced~\cite{Goerigk2017} and is now employed to evaluate nearly all modern density functionals~\cite{Goerigk2017,mehta2018semi,najibi2018nonlocal}. That same year, the even larger Main Group Chemistry Database (MGCDB84) was also presented, providing a comprehensive benchmark of over 200 density functionals~\cite{mardirossian2017thirty}. Other significant efforts were also reported.\cite{haoyu2015nonseparable,liu2023supervised}

Given that these state-of-the-art databases are now about eight years old, there is a significant opportunity to improve diversity and quality of data in new compilations. This work reports the result of our effort to develop a larger and more comprehensive database, Gold-Standard Chemical Database 137 (GSCDB137). GSCDB137 aims to deliver gold-standard accuracy and contains 137 datasets encompassing a total of 8377 individual data points (requiring around 14k single-point energy calculations). Key improvements in diversity include extensive transition-metal data drawn from realistic organometallic reactions~\cite{iron2019evaluating,Maurer_2021,dohm2018comprehensive} and well-defined model complexes~\cite{Chan_2019,chan2023dapd}. A second important addition is energy differences that reflect molecular properties. Recently, there has been controversy regarding the accuracy of electron densities from modern functionals~\cite{medvedev2017density} and thus we add density-dependent properties, such as dipole moments~\cite{hait2018dipole}, polarizabilities~\cite{hait2018polarizability,Zhu_2023}, and the field-dependence of energies~\cite{scheele2023investigating}. Additionally, we introduce datasets for vibrational frequencies~\cite{hoja2024v30}, further extending the range of available benchmarking targets. Beyond these brand-new energy categories, we also expand the coverage and diversity within established categories. Furthermore, we carefully and systematically prune all questionable data points, especially those potentially affected by spin-symmetry-breaking~\cite{lee2018regularized,lee2019distinguishing,liu2025revisiting}, yielding a benchmark suite aimed at gold-standard accuracy across an unprecedented range of chemistry.

The outline of the paper is as follows. In the next section, we provide an overview of the Gold-Standard Chemical Database 137 (GSCDB137), including a comprehensive description of all 137 benchmark sets, their characteristics, and reference values (Table~\ref{chap5tab:overview}). In Section~\ref{chap5sec:consideration}, we discuss important considerations in constructing the database and choosing reference values. This includes the integration of existing benchmark databases (MGCDB84 and GMTKN55), updates and adjustments to reference values, resolution of spin contamination issues, inclusion of newly added data sets---particularly for transition metals and molecular properties---and basis set considerations. In Section~\ref{chap5sec:bench}, we present a DFT benchmark study that evaluates the performance of 29 selected density functionals across the new database, identifying the strengths and limitations of the functionals themselves and their corresponding functional design. Finally, in Section~\ref{chap5sec:conclusion}, we present our conclusions, summarizing the key findings and discussing the implications for future density functional development.

\section{Overview of the Database}

\begin{figure}[!ht]
    \centering
    \includegraphics[width=\textwidth]{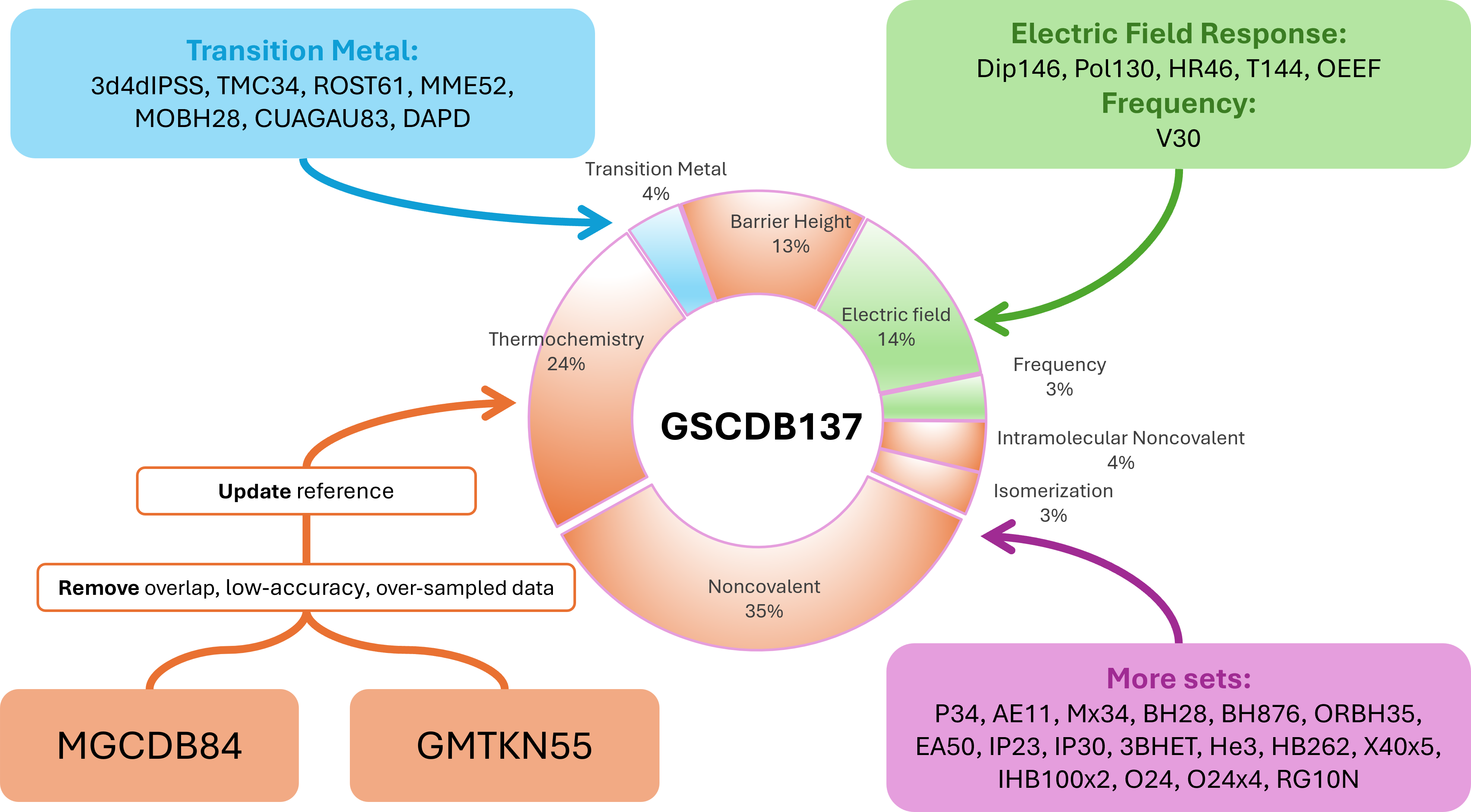}
    \caption{Overview of GSCDB138 Database generation and composition.}
    \label{fig:summary}
\end{figure}

Figure~\ref{fig:summary} summarizes the generation and composition of GSCDB137. Table~\ref{chap5tab:overview} is a simple description of all 137 benchmark sets, including a short description, the number of data points contained in each set, the root-mean-square value of their energy differences, and the reference. A more detailed version is attached in the Supplemental Information.

\footnotesize
\begin{longtable}{p{2.2cm}p{7.5cm}rp{1.1cm}p{1.1cm}p{1.3cm}}
\caption{Overview of the Gold-Standard Chemical Database 137 (GSCDB137)}\label{chap5tab:overview}\\
\hline
Name  & Description & $\#$ & RMS$\Delta E$ & Ref value & Ref data set \\
\hline
\endfirsthead

\hline
\multicolumn{6}{|c|}{Continuation of Table \ref{chap5tab:overview}}\\
\hline
Name   & Description & $\#$ & RMS$\Delta E$ & Ref value & Ref data set \\
\hline
\endhead

\hline
\endfoot

\hline
\multicolumn{6}{|c|}{End of Table}\\
\hline
\endlastfoot
\multicolumn{6}{l}{Barrier Height}\\
BH28 & Highly accurate subset chosen from BHPERI, CRBH20, BHDIV10, and PX13 sets & 28 & 35.18 & \cite{Karton_2019} &  \\
BH46 & Barrier heights of hydrogen transfer, heavy atom transfer, nucleophilic substitution, unimolecular and association reactions. This set comprises the remaining data points from BH76 after excluding those included in DBH22. & 46 & 25.14 & \cite{Goerigk2017} & \cite{zhao2005multi,zhao2005benchmark} \\
BH876 & Comprehensive reaction barrier heights & 876 & 26.93 & \cite{Prasad_2022} &  \\
BHDIV7 & Diverse reaction barrier heights & 7 & 48.77 & \cite{Goerigk2017} & \\
BHPERI11 & Barrier heights of pericyclic reactions & 11 & 23.00 & \cite{Karton_2015Accurate} & \cite{goerigk2010general} \\
BHROT27 & Barrier heights for rotation around single bonds & 27 & 8.06 & \cite{Goerigk2017} & \\
CRBH14 & Barrier heights for cycloreversion of heterocyclic rings  & 14 & 45.85 & \cite{yu2015reaction} &  \\
DBH22 & Highly accurate subset of BH76  & 22 & 29.22 & \cite{Karton_2008} & \cite{Zheng_2007} \\
INV23 & Inversion/Racemization barrier heights & 23 & 36.57 & \cite{Goerigk_2016} &  \\
ORBH35 & Difficult barrier heights for oxygen reactions (e.g., vinylperoxy radical and CO3) & 35 & 32.86 & \cite{Chan_2018} & \\
PX9 & Proton-exchange barriers in H2O, NH3, and HF clusters & 9 & 38.87 & \cite{Karton_2012Determination} & \\
WCPT26 & Barrier heights of water-catalyzed proton-transfer reactions & 26 & 36.11 & \cite{Karton_2012Assessment} &  \\[1.5cm]
\multicolumn{6}{l}{Electric Field Property and Frequency}\\
Dip146 & Dipole moments for 152 small systems & 190 & 0.12 & \cite{hait2018dipole} & \\
HR46 & Static polarizabilities for 46 systems & 128 & 0.47 & \cite{Zhu_2023} & \cite{hickey2014benchmarking} \\
OEEF & Relative energies in oriented external electric fields compared to zero field & 128 & 18.07 & \cite{scheele2023investigating} & \\
Pol130 & Static polarizabilities for 132 small systems & 296 & 1.64 & \cite{hait2018polarizability,brakestad2020static} & \\
T144 & Static polarizabilities for 144 systems & 421 & 0.69 & \cite{Zhu_2023} & \cite{wu2015choosing} \\
V30 & Frequencies of small molecular dimers with different polarity combinations (polar-polar, polar-nonpolar, and nonpolar-nonpolar) &275 & 6.0e-4 & \cite{hoja2024v30} &  \\[1.5cm]
\multicolumn{6}{l}{Isomerization Energy}\\
AlkIsomer11 & Isomerization energies of alkanes with 4-8 carbon atoms & 11 & 1.81 & \cite{Karton_2009} &  \\
C20C246 & Isomerization energies of C20 and C24 & 6 & 41.54 & \cite{Manna_2016} &  \\
C60ISO7 & Isomerization energies of C60 isomers & 7 & 102.22 & \cite{Sure_2017} &  \\
DIE60 & Isomerization energies for double-bond migration in conjugated dienes & 60 & 5.06 & \cite{Yu_2014} & \\
EIE22 & Isomerization energies of enecarbonyls & 22 & 4.97 & \cite{Yu_2015An} & \\
ISO34 & Isomerisation energies of small- and medium-sized organic molecules & 34 & 20.30 & \cite{Goerigk2017} & \cite{grimme2007compute} \\
ISOL23 & Isomerisation energies of large organic molecules & 23 & 28.03 & \cite{werner2023accurate} & \cite{huenerbein2010effects} \\
ISOMERI- ZATION20 & Isomerization energies from the W4-11 data set & 20 & 44.05 & \cite{karton2011w4} & \\
PArel & Isomerization energies of protonated isomers & 20 & 6.23 & \cite{Goerigk2017} & \\
Styrene42 & Isomerization energies of C8H8  & 42 & 68.86 & \cite{Karton_2012Explicitly} & \\
TAUT15 & Isomerization energies of tautomers & 15 & 4.52 & \cite{Goerigk2017} & \\[1.5cm]
\multicolumn{6}{l}{Intramolecular Noncovalent Interactions}\\
ACONF & Relative energies in conformers of alkane conformers  & 15 & 2.23 & \cite{Gruzman_2009} & \\
Amino20x4 & Relative energies in conformers of amino acid conformers & 80 & 2.98 & \cite{Kesharwani_2016} & \\
BUT14DIOL & Relative energies in conformers of butane-1,4-diol conformers & 64 & 2.91 & \cite{Goerigk2017} & \cite{kozuch2014conformational} \\
ICONF & Relative energies in conformers of inorganic systems & 17 & 4.54 & \cite{Goerigk2017} & \\
IDISP & Intramolecular dispersion interactions & 6 & 25.23 & \cite{Goerigk2017} & \cite{goerigk2010general} \\
MCONF & Relative energies in conformers of melatonin conformers & 51 & 5.36 & \cite{Goerigk2017} & \cite{fogueri2013melatonin} \\
PCONF21 & Relative energies in conformers of tri- and tetrapeptide conformers & 18 & 1.79 & \cite{Goerigk2017} & \cite{vreha2005structure,goerigk2013accurate} \\
Pentane13 & Relative energies in conformers of stationary points on the n-pentane torsional surface  & 13 & 6.75 & \cite{Martin_2013} &  \\
SCONF & Relative energies in conformers of sugar conformers & 17 & 4.96 & \cite{Goerigk2017} & \cite{csonka2009evaluation} \\
UPU23 & Relative energies in conformers of RNA-backbone conformers  & 23 & 6.52 & \cite{Kruse_2015} & \\[1.5cm]
\multicolumn{6}{l}{Noncovalent Interaction}\\
3B-69 & 3-Body non-additive interaction energies in three different orientations of 23 molecular crystals  & 69 & 0.69 & \cite{Rezac2015Benchmark} & \\
3BHET & 3-Body non-additive interaction energies in molecular trimers & 20 & 2.01 & \cite{Ochieng_2023} &  \\
A19Rel6 & Relative energies of 19 complexes from the A24 set on potential energy curves (PECs) & 114 & 1.65 & \cite{Witte_2015} &  \\
A24 & Highly accurate binding energies of small non-covalent complexes  & 24 & 2.65 & \cite{Rezac2015Extensions} & \cite{Rezac2013Describing} \\
ADIM6 & Interaction energies of n-alkane dimers & 6 & 3.66 & \cite{Goerigk2017} & \cite{grimme2010consistent} \\
AHB21 & Interaction energies in anion-neutral dimers  & 21 & 26.17 & \cite{Lao_2015} & \\
Bauza30 & Binding energies of halogen-, chalcogen-, and pnicogen-bonded dimers  & 30 & 23.65 & \cite{Otero_de_la_Roza_2014} & \cite{Bauza_2013} \\
BzDC215 & PECs for benzene interacting with two rare-gas atoms and eight first- and second-row hydrides  & 215 & 1.81 & \cite{Crittenden_2009} &  \\
CARBHB8 & Binding energies of hydrogen-bonded complexes between carbene analogues and H2O, NH3, or HCl  & 8 & 7.47 & \cite{Goerigk2017} & \\
CHB6 & Interaction energies in cation-neutral dimers  & 6 & 27.85 & \cite{Lao_2015} & \\
CT20 & Binding energies of charge-transfer complexes  & 20 & 1.07 & \cite{Steinmann_2012} &  \\
DS14 & Binding energies of complexes containing divalent sulfur & 14 & 3.70 & \cite{Mintz2012} &  \\
FmH2O10 & Binding energies of isomers of F-(H2O)10  & 10 & 168.50 & \cite{lao2015accurate} & \cite{lao2013improved} \\
H2O16Rel4 & Relative energies in conformers of (H2O)16 (boat and fused cube structures)  & 4 & 0.45 & \cite{yoo2010high} &  \\
H2O20Rel9 & Relative energies in conformers of (H2O)20 (low-energy structures)  & 9 & 2.76 & \cite{lao2015accurate} &  \cite{kazimirski2003search}\\
HB49 & Binding energies of small- and medium-sized hydrogen-bonded systems  & 49 & 14.12 & \cite{Boese_2015} &  \\
HB262 & Binding energies of hydrogen-bonded systems   & 262 & 7.2
& \cite{Rezac2020} &  \\
HCP32 & Binding energies of halogen-, chalcogen-, and pnicogen-bonded dimers  & 32 & 4.39 & \cite{Oliveira_2017} & \\
He3 & 3-Body non-additive interaction energies in helium trimers & 49 & 25.79 & \cite{lang2023three} &  \\
HEAVY28 & Binding energies between heavy element hydrides & 28 & 1.43 & \cite{Goerigk2017} & \cite{grimme2010consistent} \\
HSG & Binding energies of small ligands interacting with protein receptors & 21 & 6.63 & \cite{marshall2011basis} & \cite{faver2011formal} \\
HW30 & Binding energies of hydrocarbon-water dimers  & 30 & 2.34 & \cite{Copeland_2012} & \\
HW6Cl5 & Binding energies of Cl-(H2O)n (n = 1-6)  & 5 & 62.84 & \cite{lao2015accurate} & \cite{lao2013improved} \\
HW6F & Binding energies of F-(H2O)n (n =1-6)  & 6 & 81.42 & \cite{lao2015accurate} & \cite{lao2013improved} \\
IHB100 & Binding energies of equilibrium ionic hydrogen-bonded dimers in the HCNO chemical space. & 100 & 20.69 & \cite{Rezac2020} &  \\
IHB100x2 & Binding energies of ionic hydrogen bonds at 0.8 and 1.5 times the equilibrium distancein IHB100  & 200 & 13.87 & \cite{Rezac2020} & \\
IL16 & Interaction energies in anion-cation dimers  & 16 & 109.34 & \cite{Lao_2015} &  \\
NBC10 & PECs for BzBz (5), BzMe (1), MeMe (1), BzH2S (1), and PyPy (2) & 184 & 1.91 & \cite{marshall2011basis} & \cite{takatani2007performance,hohenstein2009effects} \\
NC11 & Binding energies of very small non-covalent complexes  & 11 & 0.48 & \cite{Smith_2014} &  \\
O24 & Interaction energies in 24 small high-spin open-shell dimers & 24 & 7.13 & \cite{madajczyk2021dataset} & \\
O24x4 & PECs for O24 & 96 & 4.20 & \cite{madajczyk2021dataset} & \\
PNICO23 & Interaction energies in pnicogen-containing dimers & 23 & 5.02 & \cite{Goerigk2017} & \cite{setiawan2015strength} \\
RG10N & PECs for the 10 rare-gas dimers involving helium through krypton  & 275 & 22.04 &  \cite{przybytek2017pair,lopez2004coupled,hellmann2008ab,meyer2015interacting,patkowski2010argon,jager2017thermophysical,hu2022ab,jager2016state} & \\
RG18 & Interaction energies in rare-gas complexes & 18 & 0.71 & \cite{Goerigk2017} & \\
S22 & Binding energies of noncovalently bound dimers & 22 & 9.65 & \cite{marshall2011basis} & \cite{jurevcka2006benchmark} \\
S66 & Binding energies of noncovalently bound dimers & 66 & 6.89 & \cite{Peter2023Pursuing} & \cite{Rezac2011_1} \\
S66Rel7 & Relative energies of 66 complexes from the S66 set on PECs & 462 & 2.74 & \cite{Santra_2022} & \cite{Rezac2011_1} \\
Shields38 & Binding energies of (H2O)n (n = 2 - 10)  & 38 & 51.61 & \cite{Manna_2017,santra2023valenceccsdtbindingwater} & \cite{Temelso_2011} \\
SW49Bind22 & Binding energies of isomers of SO42-(H2O) (n = 3 -5)  & 22 & 34.20 & \cite{Mardirossian_2013}& \\
SW49Rel28 & Relative energies in conformers of SO42-(H2O)n (n = 3-5)  & 28 & 1.55 & \cite{Mardirossian_2013} &  \\
TA13 & Binding energies of dimers involving radicals  & 13 & 22.00 & \cite{Tentscher_2013} & \\
WATER27 & Binding energies in (H2O)n, H+(H2O)n and OH-(H2O)n & 27 & 101.19 & \cite{Manna_2017,santra2023valenceccsdtbindingwater} & \cite{Bryantsev_2009} \\
X40 & Binding energies of complexes containing halogenated molecules  & 40 & 4.90 & \cite{Kesharwani_2018} & \cite{Rezac2012} \\
X40x5 & PECs for X40 & 200 & 4.02 & \cite{Kesharwani_2018} & \cite{Rezac2012} \\
XB25 & Binding energies of halogen-bonded dimers  & 25 & 5.88 & \cite{Kozuch_2013} &  \\[1.5cm]
\multicolumn{6}{l}{Thermochemistry}\\
AE11 & Absolute atomic energies of closed shell atoms from Ar to Rn & 11 & 6.5e6 & \cite{mccarthy2011accurate} & \\
AE18 & Absolute atomic energies of hydrogen through argon  & 18 & 1.5e5 & \cite{Chakravorty_1993} & \\
AL2X6 & Dimerisation energies of AlX3 compounds & 6 & 36.93 & \cite{Goerigk2017} & \cite{johnson2008delocalization} \\
ALK8 & Dissociation and other reactions of alkaline compounds & 8 & 69.97 & \cite{Goerigk2017} &  \\
AlkAtom19 & Atomization energies of n=1-8 alkane & 19 & 1829.31 & \cite{Karton_2009} &  \\
ALKBDE10 & Dissociation energies in group-1 and -2 diatomics & 10 & 105.53 & \cite{Yu_2015Components} & \\
AlkIsod14 & Isodesmic reaction energies in n = 3–8 alkanes. & 14 & 10.35 & \cite{Karton_2009} &  \\
BDE99MR & Multi-reference (MR) bond dissociation energies from W4-11 & 16 & 54.51 & \cite{karton2011w4} & \\
BDE99nonMR & Single-reference (SR) bond dissociation energies from W4-11 & 83 & 114.98 & \cite{karton2011w4} &  \\
BH76RC & Reaction energies of the BH76 set & 30 & 30.36 & \cite{Goerigk2017} & \cite{zhao2005multi,zhao2005benchmark} \\
BSR36 & Hydrocarbon bond separation reaction energies & 36 & 19.45 & \cite{Goerigk2017} & \cite{steinmann2009unified} \\
CR20 & Cycloreversion reaction energies & 20 & 22.31 & \cite{yu2016can} & \\
DARC & Reaction energies of Diels-Alder reactions & 14 & 34.94 & \cite{Goerigk2017} & \cite{johnson2008delocalization} \\
DC13 & 13 difficult cases for DFT methods & 13 & 71.32 & \cite{Goerigk2017} & \\
DIPCS9 & Double-ionisation potentials of closed-shell systems & 9 & 650.43 & \cite{Goerigk2017} & \\
EA50 & Vertical electron affinities & 50 & 45.42 & \cite{Ermis_2021} &  \\
FH51 & Reaction energies in various (in-)organic systems & 51 & 45.35 & \cite{Friedrich_2015} & \cite{friedrich2013incremental} \\
G21EA & Adiabatic electron affinities & 25 & 41.05 & \cite{parthiban2001assessment,boese2004w3} & \cite{curtiss1991gaussian} \\
G21IP & Adiabatic ionization potentials & 36 & 265.65 & \cite{parthiban2001assessment,boese2004w3} & \cite{curtiss1991gaussian} \\
G2RC24 & Reaction energies of systems from G2/97 set & 24 & 71.05 & \cite{Goerigk2017} & \cite{curtiss1991gaussian} \\
HAT707MR & Heavy-atom transfer energies (MR) from W4-11 & 202 & 83.41 & \cite{karton2011w4} & \\
HAT707nonMR & Heavy-atom transfer energies (SR) from W4-11 & 505 & 74.79 & \cite{karton2011w4} &  \\
HEAVYSB11 & Dissociation energies in heavy-element (i.e. Ge Se Sb) compounds & 11 & 58.56 & \cite{Goerigk2017} &  \\
HNBrBDE18 & Homolytic N-Br bond dissociation energies & 18 & 56.95 & \cite{o2016dataset} & \\
IP23 & Vertical ionization potentials & 23 & 307.03 & \cite{marie2024reference} &  \\
IP30 & Vertical ionization potentials & 30 & 281.77 & \cite{ranasinghe2019vertical} &  \\
MB08-165 & Decomposition energies of 165 artificially constructed molecules, each containing 8 atoms & 165 & 154.46 & \cite{korth2009mindless} &  \\
MB16-43 & Decomposition energies of 43 artificially constructed molecules, each containing 16 atoms & 43 & 539.03 & \cite{Goerigk2017} &  \\
MX34 & Electronic atomization energies for a range of ionic clusters with the NaCl structure & 34 & 833.23 & \cite{Chan_2023} &  \\
NBPRC & Oligomerizations and H2 fragmentations of NH3/BH3, H2 activation reactions with PH3/BH3 systems & 12 & 30.91 & \cite{Goerigk2017} & \cite{goerigk2011efficient} \\
P34AE & The atomization energies of systems containing heavy p-block elements up to xenon & 44 & 617.89 & \cite{Chan_2021} & \\
P34EA & The electron affinities of systems containing heavy p-block elements up to xenon & 9 & 44.28 & \cite{Chan_2021} & \\
P34IP & The ionization potentials of systems containing heavy p-block elements up to xenon & 15 & 224.49 & \cite{Chan_2021} & \\
PA26 & Adiabatic proton affinities including amino acids & 26 & 191.81 & \cite{Goerigk2017} &  \\
PlatonicRE18 & Reaction energies of homodesmotic, isodesmic, and isogyric reactions involving platonic hydrocarbon cages, CnHn (n = 4, 6, 8, 10, 12, 20) & 18 & 227.24 & \cite{Karton_2016Heats} & \\
PlatonicTAE6 & Total atomization energies of platonic hydrocarbon cages, CnHn (n = 4, 6, 8, 10, 12, 20) & 6 & 2539.27 & \cite{Karton_2016Heats} & \\
RC21 & Fragmentations and rearrangements in organic radical cations & 21 & 43.94 & \cite{Goerigk2017} & \cite{grimme2013towards}\\
RSE43 & Radical stabilization energies & 43 & 9.96 & \cite{Goerigk2017} & \cite{neese2009assessment} \\
SIE4x4 & Self-interaction-error related problems & 16 & 38.07 & \cite{Goerigk2017} & \\
SN13 & Nucleophilic substitution energies  & 13 & 25.67 & \cite{karton2011w4} &  \\
TAE\_W4-17MR & Total atomization energies (MR) from W4-17 & 17 & 141.49 & \cite{karton2017w4} &  \\
TAE\_W4-17nonMR & Total atomization energies (SR) from W4-17  & 183 & 518.60 & \cite{karton2017w4} &  \\
WCPT6 & Tautomerization energies for water-catalyzed proton-transfer reactions  & 6 & 7.53 & \cite{Karton_2012Assessment} & \\
YBDE18 & Bond-dissociation energies in ylides & 18 & 53.11 & \cite{Goerigk2017} & \cite{zhao2012benchmark} \\[1.5cm]
\multicolumn{6}{l}{Transition Metal}\\
3d4dIPSS & Spin-state relative energies and ionization potentials for first-row (3d) and second-row (4d) transition metals & 32 & 138.97 & \cite{balabanov2006basis,figgen2008energy,luo2012evenly,luo2014density} &  \\
CUAGAU83 & Atomization, ionization, isomerization, and binding energies  of molecular systems containing Cu, Ag, and Au & 83 & 114.12 & \cite{Chan_2019,chan2021assessment} &  \\
DAPD & Difficult bonding energies of Pd-containing diatomic molecules & 12 & 64.93 & \cite{chan2023dapd} & \\
MME52 & Reaction energies and barrier heights of metalloenzyme models & 52 & 21.71 & \cite{Wappett_2023} & \cite{Rezac2011_1} \\
MOBH28 & Barrier heights involving organometallic complexes compiled from real-life organometallic catalytic problems & 56 & 26.91 & \cite{semidalas2022mobh35} & \cite{iron2019evaluating} \\
ROST61 & Reaction energies of reactions involving open-shell single-reference transition-metal complexes & 61 & 64.10 & \cite{Maurer_2021} &  \\
TMD10 & Statistically significant subsets from TMD60, which contain bonding energies for diatomic transition-metal species & 10 & 88.94 & \cite{moltved2018chemical} & \cite{chan2019assessment} \\
MOR13 & Statistically significant subsets from MOR41, which contain  the reaction energies of closed-shell organometallic reactions & 13 & 33.90 & \cite{dohm2018comprehensive} & \cite{chan2019assessment} \\
TMB11 & Statistically significant subsets from TMB50, which contain reaction barriers involving transition metal complexes & 11 & 16.14 & \cite{kang2012accurate,sun2013performance,sun2014performance,hu2015assessment} & \cite{chan2019assessment} \\
\end{longtable}
\normalsize

\section{Considerations for Database Construction} \label{chap5sec:consideration}

We summarize below several key considerations in constructing the database and selecting appropriate reference values.

\subsection{Integration of MGCDB84 and GMTKN55}

Since MGCDB84 incorporates many data sets from GMTKN30, overlaps between MGCDB84 and GMTKN55 need to be carefully identified and resolved.  The following criteria were applied to determine which data sets to retain.

\begin{itemize}
\item When one data set was determined to be more accurate, the more reliable version was retained and the overlapping counterpart excluded.
    \begin{itemize}
        \item Retained: NBPRC, BSR36 (GMTKN55); Excluded: Same sets in MGCDB84.
        \item Retained: But14diol, MCONF (GMTKN55); Excluded: Butanediol165, \newline Melatonin52 (MGCDB84).
        \item Retained: BH76RC and BH76 (BH46 and DBH22) (GMTKN55); Excluded: BH76RC, HTBH38, NHTBH38 (MGCDB84).
        \item Retained: WATER27 (GMTKN55); Excluded: H2O20Bind4, H2O20Rel4, \newline WATER27 (MGCDB84).
    \end{itemize}
\item When one set overlaps another, we retain the smaller set if we feel it serves as a representative. Otherwise, the larger set is selected.
    \begin{itemize}
        \item Retained: PX13 (GMTKN55); Excluded: PX13 and CE20 (MGCDB84).
        \item Retained: WCPT6, WCPT27 (MGCDB84); Excluded: WCPT18 (GMTKN55).
        \item Retained: X40, XB51, and XB18 (XB25) (MGCDB84); Excluded: HAL59 (GMTKN55).
        \item Retained: Amino20x4 (GMTKN55); Excluded: CYCONF and YMPJ519 \newline (MGCDB84).
        \item Retained: Shields38 (MGCDB84); Excluded: H2O6Bind8 (MGCDB84).
        \item Retained: DIE60 (MGCDB84); Excluded: CDIE20 (GMTKN55).
    \end{itemize}
\item When data sets are identical, the GMTKN55 versions are retained, including ACONF, \newline BHPERI, AHB21, CHB6, and IL16. 
\end{itemize}

\subsection{Reference Updates and Data set adjustments}

Since the release of MGCDB84 and GMTKN55, many of their constituent datasets have been expanded or updated with more accurate theoretical reference values. Additionally, some data points in these sets already had more accurate reference values available prior to their inclusion in those databases. To ensure that every data point reflects the most reliable data available, the following updates and adjustments have been made.

\begin{itemize}
    \item All identity reactions (with zero reaction energy) in MGCDB84 are removed.
    \item Only one of the duplicate reactions is kept (the one with the more accurate reference value).
    \item Some data sets are updated with improved reference values or extended data points:
    \begin{itemize}
        \item The new W4-17 set replaces W4-11 (from GMTKN55) and TAE140 (from MGCDB84) and is split into MR and nonMR subsets following the MGCDB84 convention.
        \item ISOL23, S22, and S66 are updated with the latest reference values.
        \item All values in Shields38 (BEGDB) and WATER27 are updated with the latest reference values from Ref.\citenum{Manna_2017}. Where available, additional core-valence and beyond-CCSD(T) corrections from Ref.\citenum{santra2023valenceccsdtbindingwater} are also applied. 
        \item G21IP and G21EA are updated with theoretical reference values calculated at the W3 level.
        \item The MP2-F12/VTZ-F12+HLC/VTZ values in C20C24 are replaced by MP2-F12/V\{D,T\}Z-F12+HLC/VTZ values (or MP2/V\{Q,5\}Z+HLC/VTZ, if available), as reported in the original paper.
        \item The CCSD(T)/cc-pVTZ values in Pentane14 are replaced with CCSD(T)-F12b/cc-pVTZ-F12 values from the original literature.
        \item The CCSDTQ correction from the original paper is added to NC15.
        \item Reference values for X40 are updated using interaction energies at equilibrium geometries from the recent X40x10 data set.
    \end{itemize}
    \item Several data sets or their subsets have been replaced entirely with more accurate alternatives:
    \begin{itemize}
        \item The original RG10 set in MGCDB84 is based on the Tang–Toennies potential model, which becomes inaccurate at short interatomic distances. We have compiled a new set, RG10N, using highly accurate CCSDT(Q) reference values for He$_2$, Ne$_2$, Ar$_2$, Kr$_2$, Kr-Ne, Kr-Ar, and Kr-Xe dimers, and CCSD(T) values for He-Ne, He-Ar, and Ne-Ar dimers.
        \item Selected reactions from BHPERI (reactions 1, 2, 4-7, 11-19 for CADBH), CRBH (reactions 1, 2, 10-12, 20), BHDIV10 (reactions 2, 8, 9), and PX13 (reactions 3, 6, 11, 12) form a new benchmark set, BH28 with updated, more accurate theoretical values. The corresponding reactions are excluded from the original data sets.
        \item  DBH24 (DBH22 in this database), a subset of BH76, known for superior accuracy, has been retained separately with improved reference values. The corresponding data points (Reactions 1 2 5 6 11 12 19 20 23 24 25 26 29 30 33 34 37 38 45 46 61 62 63 64) are excluded from the original BH76 set.
    \end{itemize}
    \item Some data sets are restructured to avoid overlap with other sets:
    \begin{itemize}
        \item A21x12 is down-sampled to include only 6 relative distances (0.9, 1.1, 1.2, 1.4, 1.6, and 2.0 times the equilibrium distance) across 19 systems. Interaction energies are converted to relative energies to avoid conflict with the more accurate A24 set. The resulting set is named A19Rel6.
        \item S66x8 is updated similarly, with interaction energies converted to relative values to minimize overlap with the new S66 set. The resulting set is designated S66Rel7.
        \item The original 3B-69-TRIM set describes the trimer interaction energies without explicitly excluding two-body contributions. It is replaced by a new 3B-69 dataset that only captures non-additive three-body energies. The original 3B-69-DIM and 3B-69-TRIM sets from MGCDB84 are removed.
        \item Eight duplicated data points in SW49Bind345 are removed because they are already included in SW49Rel345.
        \item XB51 (20 data points) and XB18 (8 data points) from MGCDB84 are merged, and 3 overlapping data points are removed. The resulting set is designated XB25.
        \item PlatonicHD6, PlatonicID6, and PlatonicIG6 are merged into a single set, PlatonicRE18, as their individual performance is of limited interest in the context of such a large database.
    \end{itemize}
    \item Some data sets are removed entirely due to the insufficient accuracy of their reference values. These include 3B-69-DIM, AlkBind12, CO2Nitrogen16, SW49Bind6, SW49Rel6, H2O20Bind10, and HB15.
    \item Whenever data points are removed from a dataset,  the set is renamed to reflect the change, typically by appending a number indicating the number of data points of the new set. For example, `C60ISO' becomes `C60ISO7'.
\end{itemize}

The geometry files of MGCDB84 and GMTKN55 were downloaded from the Github repository of ACCDB~\cite{morgante2019accdb}, and we fixed some typos.

\subsection{Newly Added Data Sets}

MGCDB84 and GMTKN55 only involve the energetics of main-group element systems, lacking coverage of transition metal systems and various molecular properties. To address these limitations, the GSCDB137 database includes new data sets covering transition metals, main-group metal clusters, electric response properties, vibrational frequencies, and three-body noncovalent interactions. Additionally, new data sets are incorporated to enhance coverage of electron affinities, ionization potentials, and reaction barrier heights. Note that for some data sets, the original papers do not report non-relativistic energies or individual relativistic corrections, making it infeasible to exclude relativistic contributions. Considering that the relativistic effect is usually small for light main-group elements and can be partially and implicitly described for heavy elements by the use of the PP basis sets, these data sets are retained but need to be used cautiously.

We incorporate nine new data sets containing transition metal elements:

\begin{enumerate}
    \item \textbf{3d4dIPSS} includes spin states and ionization potentials (IP) for first-row (3d) and second-row (4d) transition metal atoms. For each spin multiplicity, only  the lowest-energy state is retained. Theoretical values are adopted from Refs.~\citenum{balabanov2006basis,figgen2008energy} when available; otherwise, experimental data from Refs.~\citenum{luo2012evenly,luo2014density} are used after subtracting the Douglas-Kroll-Hess (DKH) correction calculated at the CCSD(T)/aug-cc-pwCVQZ level. Note that many spin multiplicities in Ref.~\citenum{luo2012evenly} (4dIPSS set) are incorrect, and we have corrected them in our new 3d4dIPSS set.
    \item \textbf{TMD10}, \textbf{MOR13}, and \textbf{TMB11} are statistically significant subsets of the MOR41, TMD60, and TMB50 sets, which describe the reaction energies of closed-shell organometallic reactions, bonding energies for diatomic transition-metal species, and reaction barrier heights involving transition metal complexes, respectively.
    \item \textbf{ROST61} contains reaction energies for 61 open-shell, single-reference transition metal complex reactions.
    \item \textbf{MME55} contains metalloenzyme model reaction energies and barrier heights. After excluding spin-contaminated species, the revised set contains 52 reactions (MME52).
    \item \textbf{MOBH28} includes 28 forward and 28 reverse barrier heights for organometallic complexes.
    \item CUAGAU contains reactions involving Cu, Ag, and Au atoms. Reference energies are obtained using the CM1 (reactions 8-14), CM2 (reactions 4-7), and CM3 (reactions 1-3) protocols. CUAGAU2 extends CUAGAU to include larger systems. Reference energies are calculated using W1X-G0 theory. After removing spin-contaminated species, the two sets are put together to form the \textbf{CUAGAU83} set.
    \item \textbf{DAPD} includes bonding energies of Pd-containing diatomic molecules and serves as a difficult test set.
\end{enumerate}

Seven additional data sets are added to capture molecular properties and responses under external electric fields. Because the database may be used for training new functionals in the future, all properties are computed via finite difference (FD), which can be interpreted as a form of energy difference.

\begin{enumerate}
\item \textbf{Dip146} includes dipole moments (including directional components) for 146 small molecules at equilibrium geometries. The FD step size is set to $10^{-4}$ a.u., the same as the original paper.
\item \textbf{Pol130} includes static polarizabilities (with directional components) for 130 small molecules. The FD step size for Na is set to 0.003 a.u. to reliably obtain its large value; all other species' step sizes match the values reported in the original paper. We validated FD results against analytical static polarizabilities in Ref.~\citenum{hait2018polarizability} and removed the data points with large FD errors.
\item \textbf{HR46} and \textbf{T144} include static polarizabilities for medium-sized molecules. A uniform FD step size of 0.004 a.u. is used, as in the original reference.
\item \textbf{OEEF} reports changes in electronic energies under oriented external electric fields relative to the field-free condition. The reference data are calculated at the CCSD(T)/aug-cc-pVQZ level. We observed that $\omega$B97M-V/aug-pc-3 exhibits large errors for several cases due to artificial spin-symmetry breaking under very strong fields. Notably, this issue disappears when using the def2-QZVPPD basis. The three affected difficult data points are separated into a dedicated \textbf{OEEFD} set for future investigation and are not included in the GSCDB137 benchmark.
\item \textbf{V30} contains the vibrational frequencies for molecular dimers across different polarity combinations (polar–polar, polar–nonpolar, and nonpolar–nonpolar). Reference frequencies are calculated using atomic masses of the most abundant isotopes. For numerical stability, the FD step size is adjusted based on frequency magnitude: 0.001 times the normal mode for frequencies over 1000 $\text{cm}^{-1}$, 0.003 times the normal mode for frequencies between 1000 and 100 $\text{cm}^{-1}$, and 0.01 times the normal mode for frequencies below 100 $\text{cm}^{-1}$. We validated FD results from $\omega$B97M‑V against analytical frequencies\cite{liang2023analytical} and excluded any modes where FD errors exceeded 5 $\text{cm}^{-1}$ or contributed more than 25\% of the functional error. As a result, the final set contains no vibrational modes below 40 $\text{cm}^{-1}$.
\end{enumerate}

Finally, we add the following new data sets to expand the coverage of underrepresented interactions:

\begin{enumerate}
\item \textbf{MX34}, derived from MX35 by removing spin-contaminated species, contains atomization energies (AEs) of ionic clusters resembling NaCl crystal structures. (The scalar-relativistic correction is not excluded.)
\item \textbf{P34} includes AE, IP, and electron affinities (EAs) of systems containing heavy p-block elements up to xenon, further increasing the element coverage of our new database. (The scalar-relativistic correction is not excluded.) We split this set into P34AE, P34IP, and P34EA. 
\item \textbf{AE11} provides all-electron non-relativistic atomic energies for eleven closed-shell atoms from Ar to Rn, calculated at the CCSD(T) level. It represents an extension of the AE18 set and serves as an overfitting diagnostic.
\item \textbf{EA50} contains vertical electron affinities derived by removing spin-contaminated systems and overlapping species from G21EA. It uses EKT-CCSD(T)/aug-cc-pVQZ as the reference method and replaces the EA13 set, which uses experimental reference values.
\item \textbf{IP23} includes highly accurate vertical ionization potentials of small molecules, replacing the IP13 set, which uses experimental reference values.
\item \textbf{IP30} contains vertical ionization potentials, assembled by selecting the most accurate data points from Ref.~\citenum{ranasinghe2019vertical} and removing overlaps with IP23.
\item BH9 contains forward and reverse barrier heights (BHs) of 449 diverse reactions, expanding the coverage of barrier heights. After removing spin-contaminated species and duplicate reactions, 876 BHs remain, forming the \textbf{BH876} set.
\item \textbf{ORBH35}, drawn from the Barriometry database, contains barrier heights for oxygen reactions (i.e., vinylperoxy radical and CO\textsubscript{3}), representing very difficult barrier heights. (The scalar-relativistic correction is not excluded.)
\item \textbf{3BHET} includes non-additive three-body interaction energies of small molecular trimers.
\item \textbf{He3} includes non-additive three-body dispersion-dominated interaction energies of helium trimers. Considering DFT's accuracy limitation, only data points with three-body energies exceeding 0.1 mHartree in absolute value are included.
\item \textbf{HB262} includes 262 hydrogen-bonded complexes derived from HB375~~\cite{Rezac2020}, after excluding those classified as having "no hydrogen bond."
\item \textbf{X40x5} is a distance-sampled version of X40x10, including interaction energies at 0.8, 0.85, 0.90, 0.95, 1.0, and 1.5 times the equilibrium distance.
\item IHB100x10 covers ionic hydrogen bonds across separation distances in the HCNO chemical space. We adopt data points at 0.8, 1.0, and 1.5 times the equilibrium distance. Equilibrium points form the \textbf{IHB100} set; the remainder form \textbf{IHB100x2}.
\item O24x5 provides interaction energies for 24 small high-spin open-shell dimers. Equilibrium geometries form the \textbf{O24} set; remaining distances form the \textbf{O24x4} set.
\end{enumerate}

\subsection{Spin Contamination Resolution}\label{chap5subsec:spin}

To ensure the reliability of the reference values, we address potential issues related to spin contamination. We begin by performing internal stability analyses~\cite{seeger1977self} using the $\omega$B97X-V functional with a small basis set to determine whether the SCF procedure converges to the true lowest-energy solution in the first run of the unrestricted SCF calculation. Molecules that fail this test are labeled as “Difficult.” 

These “Difficult” molecules often exhibit spin contamination. To evaluate whether spin symmetry breaking is physically justified (i.e., essential to describe part of multi-reference effects), we apply the $\kappa$-OOMP2 method with $\kappa=1.45$~\cite{lee2018regularized,liu2025revisiting}. Molecules showing essential spin symmetry breaking are identified by comparing the expected and computed $\langle S^2 \rangle$ values, as shown in Table~S1.

For molecules with essential spin symmetry breaking, we further assess the reliability of the reference data. Reference values are accepted as reliable if they are derived from experimental measurements, unrestricted CCSD(T), or beyond-CCSD(T) methods. If none of these conditions are met, the reference value is considered unreliable, and reactions involving such molecules are excluded from the database.

\subsection{Basis Set Considerations}

Consistent with the practice established in MGCDB84, most DFT calculations are carried out using the def2-QZVPPD basis set to minimize basis set incompleteness error, while keeping compute costs manageable. However, certain data sets require exceptions. In some cases, the reference method employs pseudopotentials or specially optimized basis sets, making it important to adopt the original computational setup to avoid additional error. In other cases, def2-QZVPPD is (i) too small to match the highly accurate reference, (ii) missing the core-correlation functions required for precise atomic energies, or (iii) unnecessarily large for very big molecules such as those in the C60ISO set. For these reasons, the data sets listed in Table \ref{tab:basis_sets} are evaluated with alternative basis sets chosen to match their original references or to strike an optimal accuracy versus cost balance. 

For practical purposes, def2-QZVPPD may still be used for most of these data sets, but substantial errors may occur for certain systems. Notably, large deviations are observed for the AE11 set and for excited He systems in O24x5 (O24\_13, O24x4\_49, O24x4\_50, O24x4\_51, O24x4\_52). For instance, the error of $\omega$B97M-V on O24\_13 decreases dramatically from –10.7 to +3.0~kcal~mol$^{-1}$ when the basis set is changed from def2-QZVPPD to d-aug-cc-pV5Z.

\begin{table}[h!]
\centering
\caption{Data Sets and Corresponding Basis Sets}
\label{tab:basis_sets}
\begin{tabular}{|l|l|}
\hline
\textbf{Data Set Name} & \textbf{Basis Set} \\ \hline
G21IP, IP23, IP30, G21EA, EA50 & aug-pc-4 \\ \hline
Dip152, HR46, Pol132, T144  & aug-pc-3 \\ \hline
AE18 & aug-cc-pCV5Z\\ \hline
RG10N, He3, excited He systems in O24x5 & d-aug-cc-pV5Z\\ \hline
OEEF,OEEFD & aug-cc-pVQZ \\ \hline
AE11 & Basis set in Ref.~\citenum{mccarthy2011accurate} \\ \hline
BH886, MME54, MOBH35,& def2-QZVPP \\ 
TMC34, ROST61  & \\ \hline
C60ISO7 & def2-TZVPD \\ \hline
3d4dIPSS & 3d: aug-cc-pwCVQZ; \\ & 4d: aug-cc-pwCVQZ-PP\\ \hline
DAPD & aug-cc-pwCVQZ-PP for Pd; cc-pVQZ for H; \\ & aug-cc-pwCVQZ for other elements\\ \hline
P34 & aug-cc-pwCVQZ-PP for In, Sn, Sb, Te, I, Xe;\\ & def2-QZVPPD for all others\\ \hline
\end{tabular}
\end{table}

\section{DFT Benchmark Result}\label{chap5sec:bench}
To evaluate the performance of modern density functionals, we conduct a comprehensive benchmark using the GSCDB137 database. 

\subsection{Functional Candidates and Evaluation Metric} \label{subsec:func_metric}
This subsection introduces the tested functionals and the evaluation metric.

A total of 29 functionals are selected, spanning all rungs of Jacob's ladder, including the local spin density approximations (LDA), generalized gradient approximations (GGAs), meta-GGAs, hybrid GGAs, hybrid meta-GGAs, and double hybrids (DH). A functional is included if it satisfies at least one of the following criteria: it is widely adopted in the quantum chemistry community, or it ranks among the top two performers in at least one energy category in previous benchmarks such as MGCDB84 or GMTKN55. There are two exceptions. One is the LDA category, for which we include only SPW92, as the MGCDB84 benchmark indicates negligible differences between different  LDA parameterizations. The other is DH, where the calculations are very expensive, and we only choose two available functionals in Q-Chem, $\omega$B97M(2) and revDSD-PBEP86-D4, as representative semi-empirical and non-empirical functionals.

The tested functionals are grouped as follows:

\begin{itemize}
    \item \textbf{Double Hybrids (DH):} revDSD-PBEP86-D4\cite{santra2021}, $\omega$B97M(2)~\cite{mardirossian2018survival}
 \item \textbf{Hybrid Meta-GGAs (HMGGA):} 
    $\omega$B97M-V~\cite{Mardirossian:2016}, 
    CF22D~\cite{liu2023supervised},
    r2SCAN0-D4~\cite{bursch2022dispersion}, 
    BMK-D3(BJ)~\cite{boese2004development}, 
    MN15-D3(BJ)~\cite{haoyu2016mn15}, 
    M06-2X-D3(0)~\cite{zhao2008m06}, 
    M08-HX-D3(0)~\cite{zhao2008exploring}, 
    PW6B95-D3(BJ)~\cite{zhao2005design}, 
    M05-2X-D3(0)~\cite{zhao2006design}
    
    \item \textbf{Hybrid GGAs (HGGA):} 
    $\omega$B97X-V~\cite{Mardirossian:2014}, 
    B3LYP-D4~\cite{Becke1988,Lee1988}, 
    CAM-B3LYP-D4~\cite{yanai2004new}, 
    PBE0-D4~\cite{perdew1996generalized}, 
    SOGGA11X-D3(BJ)~\cite{peverati2011communication}
    
    \item \textbf{Meta-GGAs (MGGA):} 
    B97M-V~\cite{Mardirossian:2015}, 
    r2SCAN-D4~\cite{furness2020accurate}, 
    MN15L-D3(0)~\cite{yu2016mn15}, 
    M06L-D4~\cite{zhao2006new}, 
    M11L-D3(0)~\cite{peverati2012m11}, 
    revTPSS-D4~\cite{perdew2009workhorse}
    
    \item \textbf{GGAs:} 
    PBE-D4~\cite{perdew1996generalized}, 
    revPBE-D4~\cite{zhang1998comment}, 
    B97-D4~\cite{grimme2006semiempirical}, 
    N12-D3(0)~\cite{peverati2012improved}, 
    OLYP-D4~\cite{handy2001left,Lee1988}, 
    BLYP-D3(BJ)~\cite{becke1988density,Lee1988,miehlich1989results}
    
    \item \textbf{LDA:} SPW92~\cite{dirac1930note,perdew1992accurate}
\end{itemize}

If a functional does not come with its own dispersion correction, we add D4~\cite{grimme2016dispersion, caldeweyher2019generally}. If no D4 parameter set is available, D3 is used instead. Dispersion energies are computed using the \texttt{dftd4}~\cite{dftd4package} and \texttt{simple-dftd3}~\cite{dftd3package} packages. All dispersion parameters are set to their default values as defined in the respective packages (note that using the recent smooth D3S and D4S versions\cite{tkachenko2024smoother,tkachenko2024smooth} of D3 and D4 is expected to make no chemically significant difference to the results\cite{tkachenko2024smoother,tkachenko2024smooth}).

All the DFT calculations are performed using a development version of Q-Chem 6.0~\cite{epifanovsky2021software}, except for revDSD-PBEP86-D4, which was evaluated with ORCA 6.1~\cite{ORCA6}. For most calculations, a (99,590) grid (99 radial shells with  590 grid points per shell) is used for semi-local functional integrals, and SG-1, a subset of (50, 194), is used for non-local VV10 correlation\cite{vydrov2010nonlocal}. A denser grid, up to (500,974) for semi-local integral and (75,302) for non-local integral, will also be used when necessary, such as for dispersion-bonded sets (RG10N, He3), for sets with very large reference values (AE11, AE18), or for some difficult molecules where some meta-GGAs strive to get SCF converged in the small grid setup.

We assess the accuracy of each density functional using the mean absolute error (MAE) computed for each data set in the GSCDB137 database, with the exception of Dip146, Pol132, HR46, T144, OEEF, OEEFD, O24, O24x4, TMD10, MOR13, and TMB11. For Pol132, HR46, T144, OEEF, and OEEFD, mean absolute relative error (MARE) is used instead to account for the wide variation in the magnitudes of the reference values. For Dip146, the mean absolute regularized error defined in the original paper is used~\cite{hait2018dipole}. For the TMD10, MOR13, and TMB11 sets, we follow the recommendation of the original publications and adopt estimated (weighted) mean absolute deviations. For O24 and O24x4, weights are assigned according to the general performance of hybrid functionals on each data point within O24. All weighting factors are provided in the supplemental file \texttt{DatasetEval.xlsx}.

To establish a robust baseline for comparison, we define a “standard error” for each data set as the average of the second, third, and fourth lowest errors among all tested hybrid functionals. Hybrid functionals are chosen for this reference because of their widespread use in chemistry. Excluding the best-performing functional helps avoid bias from potential overfitting on individual data sets.

The performance of each functional is quantified using a \textbf{normalized error ratio (NER)}, defined as the ratio between its error and the standard error for a given data set. These ratios are then averaged across all data sets within each property category---barrier heights (BH), electric field responses (EF), vibrational frequencies (FREQ), isomerization energies (ISO), noncovalent interactions (NC), intramolecular noncovalent interactions (INC), thermochemical properties (TC), and transition metal systems (TM)---as well as across the entire GSCDB137 database to yield an overall mean score for each functional.

\subsection{A Bird's-Eye View of the Benchmark Results}

\begin{figure}[!ht]
    \centering
    \includegraphics[width=\textwidth]{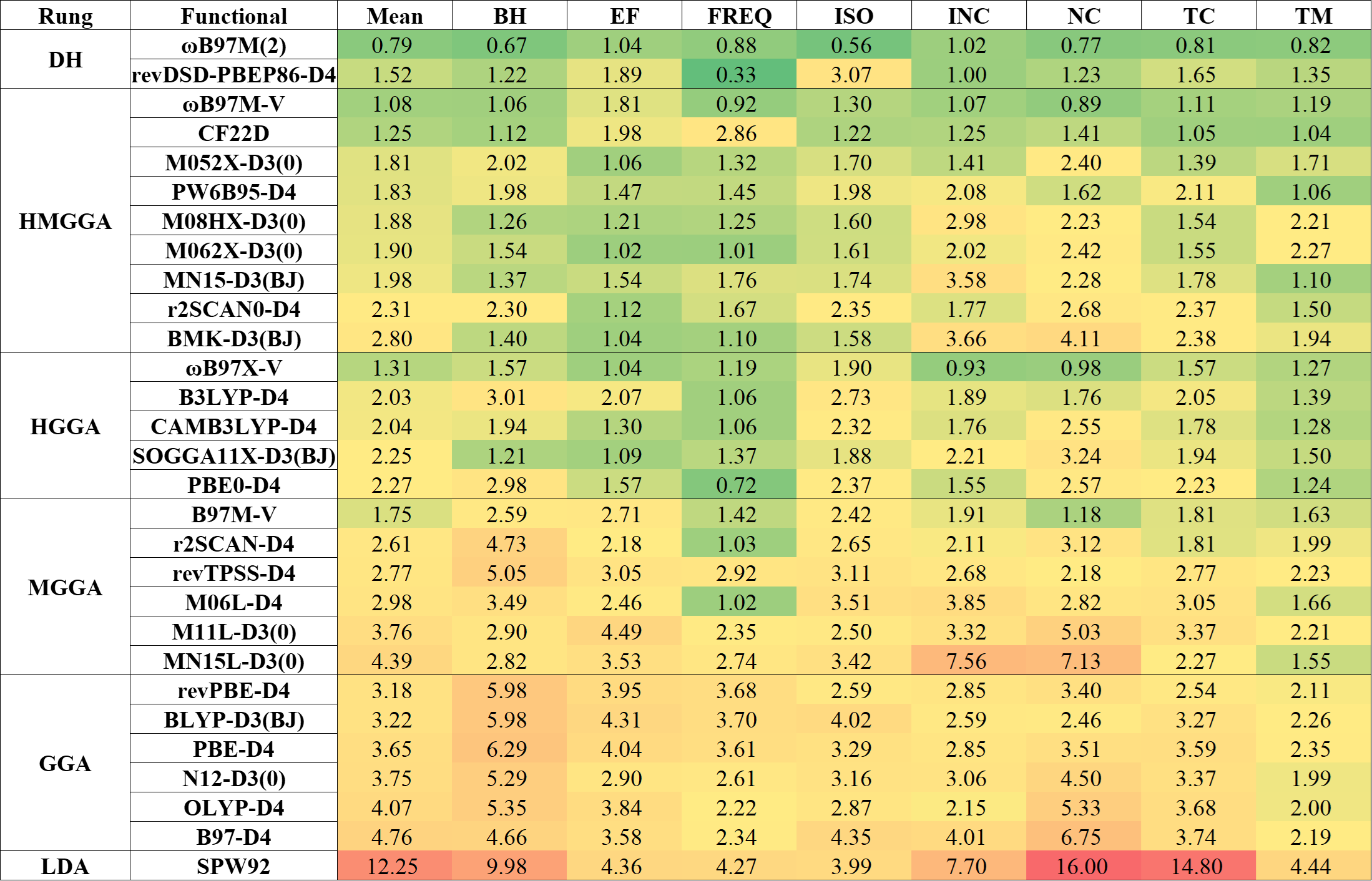}
    \caption{Normalized mean absolute error ratios (NERs) of 29 tested density functionals across seven property categories and their overall mean. Values are relative to the  baseline (average of 2nd-4th best hybrid functionals) for each data set.}
    \label{fig:Rel_heatmap}
\end{figure}

Figure~\ref{fig:Rel_heatmap} 
summarizes the benchmark results for the 29 selected density functionals across seven property categories and their overall mean performance. Overall, functional performance improves along Jacob’s ladder from LDA to double hybrids. The best double hybrid consistently yields the most accurate predictions, followed by the best HMGGA functionals and the best HGGA. In contrast, GGAs and particularly the LDA trail markedly, underscoring the benefits of exact exchange and higher-order correlation. These overall trends and relative rankings are consistent with those observed for the GMTKN55 database. However, when the new EF, FREQ, and TM categories are examined individually, additional distinctions in functional performance emerge that are not captured by GMTKN55. For EF, the best HMGGA (M062X-D3(0)) achieves only a marginally lower mean normalized error (NER) than the best HGGA, $\omega$B97X-V. Moreover, if root-mean-squared relative/regularized EF errors are used, as in Ref.~\citenum{hait2018dipole,hait2018polarizability}, $\omega$B97X-V would actually outperform M062X-D3(0). For vibrational frequencies, Jacob’s ladder generally holds, but the top performers—revDSD-PBEP86-D4 and PBE0-D4—are both nonempirical functionals. Nonetheless, hybrid functionals such as $\omega$B97M-V (except for EF) and $\omega$B97X-V still achieve strong performance, highlighting their broad transferability.

Among HMGGAs, $\omega$B97M-V delivers the lowest overall mean NER (1.08) and leads BH, FREQ, INC, and NC while remaining competitive in other categories except for EF. As shown in Figure~S1, it achieves “Good” performance (error smaller than that of the fourth-best hybrid) on 88 of the 137 sets. The second-best performer in this rung is CF22D (overall mean NER = 1.25; “Good” count = 74). CF22D is a supervised machine-learned hybrid meta-GGA with more than fifty adjustable parameters, developed using an active-learning strategy based on the MN15 functional form. It shows broad improvements over earlier Minnesota functionals such as its parent, MN15-D3(BJ), likely because its training set includes more diverse NC and ISO data. At the same time, $\omega$B97M-V has weaker EF accuracy compared to $\omega$B97X-V, and CF22D has poorer performance for EF and FREQ compared to MN15-D3(BJ). This illustrates that HMGGAs can overfit to their training sets and, as a consequence, can lose accuracy on properties outside that domain. We therefore recommend $\omega$B97M-V as the default HMGGA, with CF22D as a somewhat poorer alternative for codes that lack VV10 support. For EF-related properties, M062X-D3(0) or BMK-D3(BJ) may be preferable.

Among HGGAs, $\omega$B97X-V shows excellent and balanced performance (mean NER = 1.31; “Good” count = 69), ranking as the third-best hybrid functional overall. It performs particularly well for EF, even outperforming $\omega$B97M-V. Only SOGGA11X-D3(BJ) surpasses $\omega$B97X-V significantly in BH. Despite its historical popularity, B3LYP performs poorly across most categories and, consistent with earlier conclusions,\cite{Goerigk2017,mardirossian2017thirty} can no longer be recommended beyond geometry optimization and vibrational analysis. Our previous benchmark also indicated that a triple-zeta basis set is required for reliable frequency calculations with B3LYP.\cite{liang2023analytical} Therefore, the commonly used B3LYP/6-31G* combination should be phased out.

In the MGGA category, B97M-V delivers the best overall performance (mean NER = 1.75), with especially strong results for NC, ranking just behind $\omega$B97M-V and $\omega$B97X-V. Its performance is somewhat weaker for EF and FREQ, where r2SCAN-D4 is the best functional. The recently developed deep learning functional Skala is not tested here because its coefficients have not yet been released when our paper is submitted. However, it is expected to outperform traditional MGGA functionals, as it contains additional non-local information. GGA functionals consistently yield higher errors, with mean NER typically exceeding three. Among them, revPBE-D4 provides the most balanced performance and is recommended for general use, except for EF and FREQ, where N12-D3(0) and OLYP-D4 are superior, respectively. SPW92, the LDA functional, yields the worst overall performance (mean NER = 12.25), confirming the well-known limitations of the LDA for chemical applications.

Two further points deserve mention. First, FREQ behaves differently from other categories, raising questions about the representativeness of the single set V30. To probe this, we revisited our earlier benchmark~\cite{liang2023analytical} and confirmed that PBE0‑D3(BJ) indeed yields the lowest MAE on V30 and across the broader noncovalent frequency benchmark (within the scope of tested functionals). In contrast, B3LYP‑D3(BJ) performs best on covalent frequencies and overall. Thus, V30 alone may not fully capture the diversity of vibrational properties. However, given its reasonable coverage of intramolecular covalent interactions and the computational cost of larger frequency datasets, V30 remains the most practical single-set choice. Second, r2SCAN-D4 outperforms both revPBE-D4 and PBE-D4 across all categories. However, its hybrid counterpart, r2SCAN0-D4, shows similar mean NER to PBE0-D4 but diverges across individual categories. This suggests that improvements achieved from semi-local functional design advances do not necessarily transfer to their hybrid counterparts.

\subsection{Property-Specific Benchmark Analysis}

\begin{figure}[!ht]
    \centering
    \includegraphics[width=\textwidth]{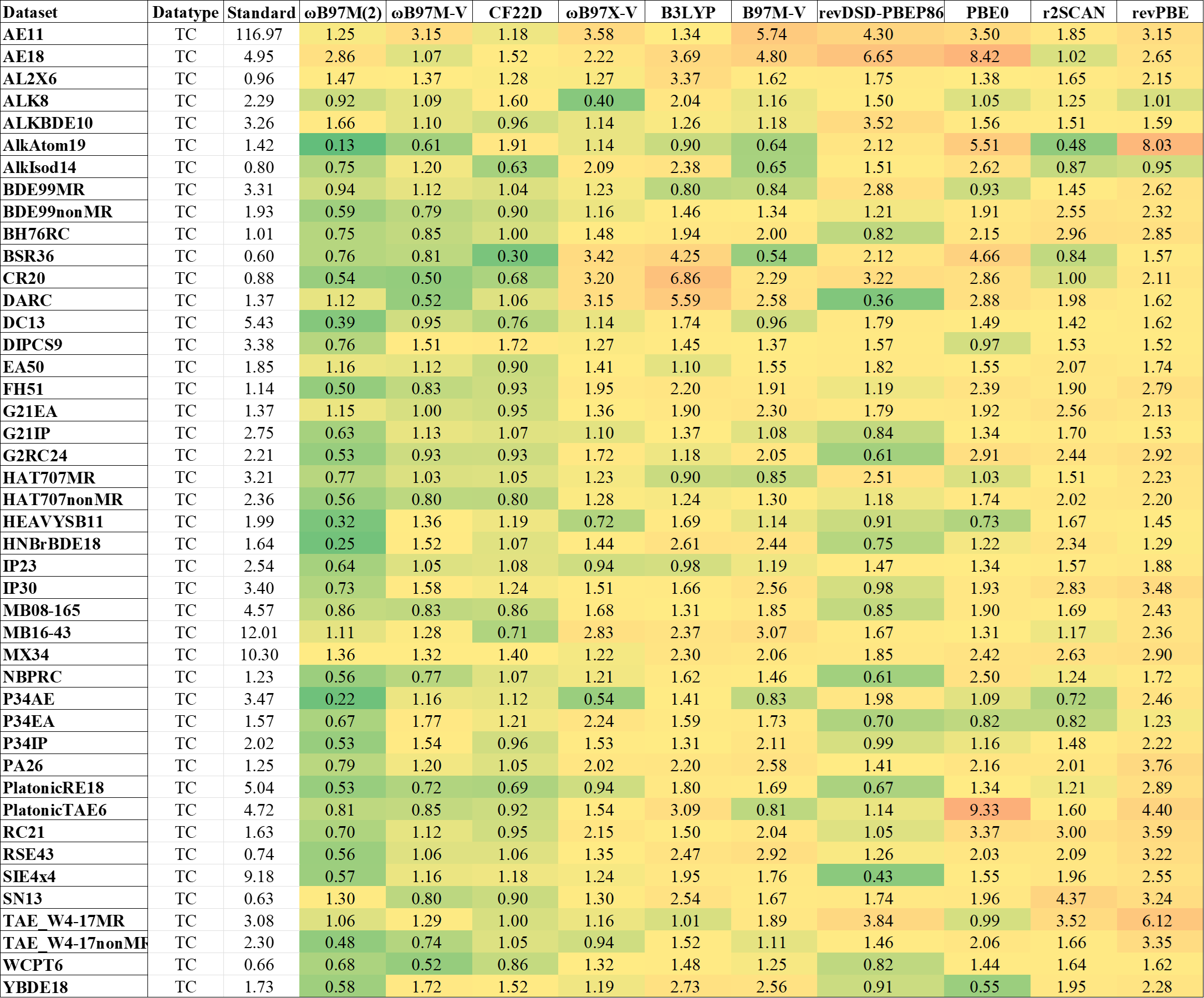}
    \caption{Normalized mean absolute error ratios (NERs) for 10 representative functionals on thermochemistry (TC) data sets. Each cell is the NER of a functional on a particular data set, relative to the hybrid baseline (“Standard error” in kcal/mol). For brevity, dispersion-correction labels are omitted.}
    \label{TC}
\end{figure}

\begin{figure}[!ht]
    \centering
    \includegraphics[width=\textwidth]{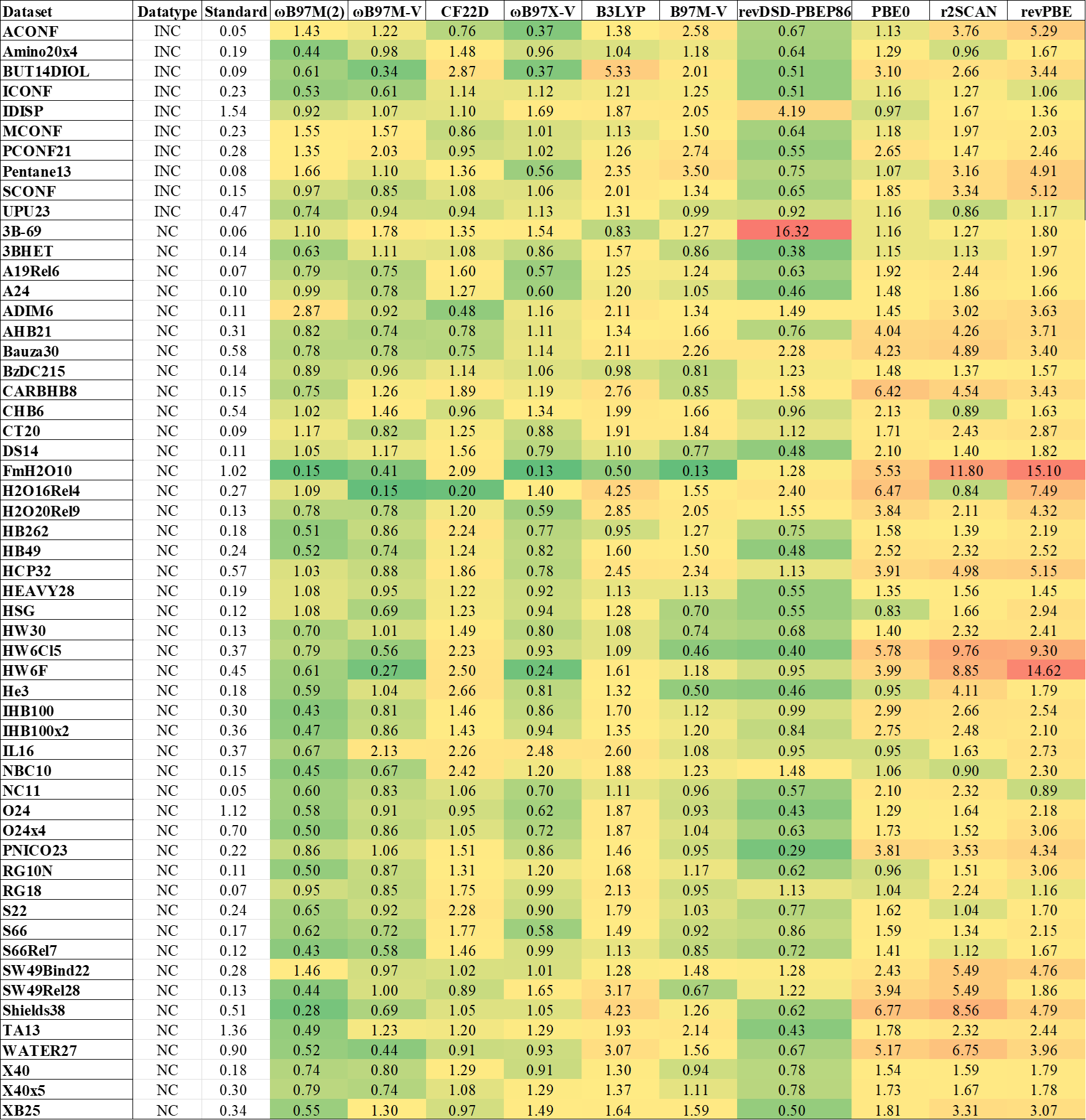}
    \caption{Normalized mean absolute error ratios (NERs) for 10 representative functionals on noncovalent interaction (NC) and intramolecular NC (INC) data sets. Each cell is the NER of a functional on a particular data set, relative to the hybrid baseline (“Standard error” in kcal/mol except for O24 and O24x4). For brevity, the names of the dispersion corrections are omitted.}
    \label{NC}
\end{figure}

\begin{figure}[!ht]
    \centering
    \includegraphics[width=\textwidth]{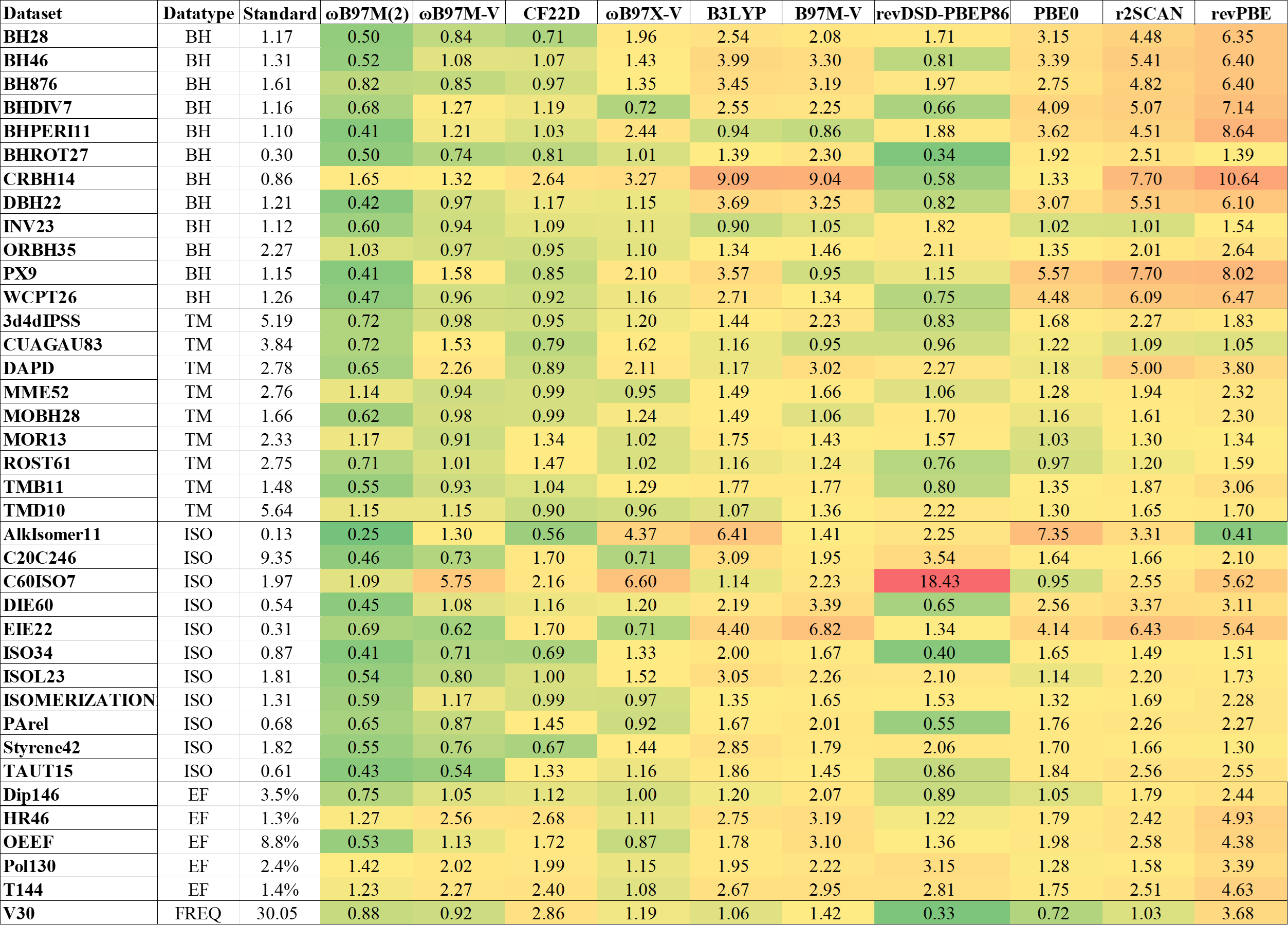}
    \caption{Normalized mean absolute error ratios (NERs) for 10 representative density functionals across the remaining categories: barrier heights (BH), transition-metal systems (TM), isomerization energy (ISO), electric-field responses (EF), and vibrational frequencies (FREQ). Each cell represents the NER of a functional on a particular data set relative to the hybrid baseline (“Standard error”), expressed in kcal~mol$^{-1}$ unless otherwise specified for the corresponding property set or TM sets (see Sec.~\ref{subsec:func_metric}). For clarity, dispersion-correction labels are omitted.}
    \label{BHetc}
\end{figure}

Figures~\ref{TC}-\ref{BHetc} present the NERs for ten representative functionals across all data sets, alongside the hybrid baseline (``Standard'' error). We begin by discussing hybrid functionals, followed by comments on double hybrids in the next subsection.

For main-group thermochemistry (TC) (Figure \ref{TC}), standard MAEs typically fall between 1 and 5 kcal/mol, indicating that even the best hybrid functionals still fail to reach chemical accuracy (1 kcal/mol) for most sets. Particularly large errors are observed for AE11 (atomic energies of heavy elements), MX34 (atomization energies of NaCl-like ionic clusters), and MB16-43 (mindless benchmarks). Ionization potential (IP) data sets, including DIPCS9 (double IP), generally show MAEs around 2--3.5 kcal/mol, while electron affinity (EA) sets average below2 kcal/mol. Among all tested functionals, CF22D and \(\omega\)B97M-V perform best, with CF22D showing clear advantages on challenging sets like MB16-43 and TAE\_W4-17MR. Notably, although \(\omega\)B97M-V and \(\omega\)B97X-V perform similarly on IP and EA, \(\omega\)B97X-V outperforms \(\omega\)B97M-V on time-dependent DFT (TDDFT) excitation energies~\cite{liang2022revisiting}, suggesting little correlation between accuracy in adiabatic IP/EA and accuracy for TDDFT excitation energies (which depends on the additional adiabatic approximation). It is possible that orbital-optimized DFT (OO-DFT) excitation energies\cite{hait2021orbital} might show stronger correlations. Furthermore, \(\omega\)B97M-V and \(\omega\)B97X-V yield reasonable performance on MX34 but greatly overestimate solid-state band gaps~\cite{lee2022faster}, implying that MX34 may not fully capture periodic system behavior (we also note that solid-state gaps are eigenvalue differences while MX34 are total energy differences).

Barrier‑height (BH) data (Figure \ref{BHetc}) tell a more comforting story: good hybrids sit between 1 and 2~kcal~mol$^{-1}$. That might look like an improvement over TC, but remember that BH reference values are usually smaller in this database, so the same absolute error can hide larger percentage mistakes. $\omega$B97M‑V and CF22D perform best in this category, though CF22D shows large errors on the cycloreversion subset CRBH14.

For transition metal (TM) systems (Figure \ref{BHetc}), top-performing functionals achieve MAEs around 1--3 kcal/mol for organometallic complexes, but significantly higher errors (\(\sim\)5 kcal/mol) occur for metal-only systems such as 3d4dIPSS, CUAGAU83, and TMD10. Performance varies with reaction type. CF22D performs poorly on reaction energies in MOR13 and ROST61, despite having the lowest overall NER for TM. In contrast, \(\omega\)B97M-V excels for organometallic reaction energies and barriers but struggles with small cluster sets like CUAGAU83 and DAPD. Interestingly, the performance gap between empirical and non-empirical functionals narrows for TM systems, suggesting reduced advantages of empirical parameterization, at least when extrapolating beyond training data.

For Isomerization (ISO) sets (Figure \ref{BHetc}), standard MAEs typically fall around 1--2 kcal/mol except for C20C246.  \(\omega\)B97M-V fails badly on C60ISO7 though it performs very well on other sets, which makes CF22D the best overall. consistent with known problems for range-separated hybrids on. As noted in the source study \cite{karton2018performance,karton2022fullerenes},  fullerene isomer energies are particularly challenging for functionals with a high percentage of long-range exact exchange. Interestingly, $\omega$B97M(2) remedies this issue, suggesting that a balanced cancellation between exchange and correlation at medium-to-long ranges is essential. It is also worth noting that EIE22 remains challenging for semi-local functionals.

For noncovalent interactions (NC) and intramolecular NC (INC) (Figure \ref{NC}), sub-kcal/mol accuracy is generally attainable except for radical systems (TA13, O24) and IDISP. \(\omega\)B97M-V, \(\omega\)B97X-V, and B97M-V lead their respective rungs, echoing findings from MGCDB84 and GMTKN55. These functionals exhibit low errors for ionic hydration (e.g., FmH2O10, HW6Cl, HW6F), possibly reflecting overfitting. Conversely, they perform poorly on the 3B-69 set, indicating that VV10 lacks the explicit three-body terms captured by DFT-D4. When such three-body terms from $\omega$B97M-D4 are included, the NERs of $\omega$B97M-V improve significantly—from 1.78 to 1.21 for 3B-69 and from 1.11 to 0.98 for 3BHET. Moreover, the mean NER for INC decreases from 1.07 to 0.81, although the mean NER for NC remains largely unchanged. We therefore recommend adding three‑body dispersion to VV10‑based DFAs for large systems.

For vibrational frequencies (Figure \ref{BHetc}), the typical MAE is around 30 $\text{cm}^{-1}$, with the best-performing hybrid and double hybrid functionals, PBE0‑D4 and revDSD-PBEP86-D4, reducing this to 22 $\text{cm}^{-1}$ and 10 $\text{cm}^{-1}$, respectively. For dipole moments, the mean absolute regularized error is about 3.5\%. Static polarizabilities show a standard MARE of 2.4\% on Pol130 but only about 1.3\% on HR46 and T144, reflecting the greater challenge of spin-polarized or exotic molecules in Pol130. For orientation-dependent electric fields (OEEF), the standard MARE rises to 8.8\%, indicating that stronger fields amplify functional errors. Interestingly, for electric field properties (EF) (Figure \ref{BHetc}), \(\omega\)B97M-V and CF22D clearly underperform other hybrids, contrary to trends in ground-state energy performance. This reversal reinforces the concern that \(\omega\)B97M-V and CF22D may overfit energy-based training sets and highlights the need to include molecular response properties during functional development to ensure broader transferability.

\subsection{Additional Considerations for Double Hybrid Functionals}

Compared to hybrid functionals, double-hybrid (DH) functionals offer substantial improvements across most data sets. The overall mean NERs of $\omega$B97M(2) and revDSD-PBEP86-D4 are 0.79 and 1.52, respectively—representing relative improvements of 27\% and 33\% compared to their hybrid analogues, $\omega$B97M-V and PBE0-D4. However, these gains come with new challenges due to the inclusion of MP2 correlation, which is a quite limited approach to wavefunction correlation. The presence of MP2 may introduces sensitivity to computational choices of certain functionals, such as basis set size, frozen-core treatment, and spin contamination (a proxy for strong correlation in many cases\cite{shee2021revealing,ganoe2024notion,liu2025revisiting}).

One important consideration is basis-set incompleteness. The MP2 component converges only as $L^{-3}$ with the maximum angular momentum $L$, whereas the SCF part approaches the CBS limit exponentially. \cite{martin2020empirical} A quadruple-\emph{zeta} basis is therefore still far from complete for a double hybrid. In practice one should use the same basis that was employed during parametrization. Accordingly, we pair $\omega$B97M(2) with def2-QZVPPD whenever possible, since that is how the functional is defined. An instructive exception is the AE11 atomic set: a specially designed core-correlated basis reduces the error much more than def2-QZVPPD, indicating that explicit CBS extrapolation or custom basis optimization can be essential for the most demanding cases. The result also suggests that future double hybrids should be trained with larger, explicitly CBS-compatible basis sets. Two promising strategies have recently been proposed to mitigate the slow basis-set convergence of double hybrids: F12 correlation treatments \cite{mehta2022explicitly} and density-based basis-set corrections \cite{mester2025near,znaida2025double}. Incorporating such approaches may enable the development of next-generation double hybrids with improved basis-set efficiency.

One more important source of error in DH-DFT the use of frozen core (FC) approximations. Since the semi-local component of a DH is fitted with a specific FC protocol, calculations should, in principle, replicate that protocol.  For example, $\omega$B97M(2) was trained with the FC approximation. If this is omitted, the MAE on the AlkAtom19 set rises dramatically from 0.18 kcal/mol to 6.24 kcal/mol—an increase by a factor of 35. However, the problem becomes more complicated when considering elements outside the training set or software-specific definitions of frozen orbitals. For instance, in Q-Chem, the K$^+$ ion is treated with 9 frozen orbitals (a very large core!) by default, leading to an 8.15 kcal/mol MAE on the CHB6 set. Without FC, the error drops to just 1.23 kcal/mol. Another extreme case is AE11, where enabling FC inflates the error by more than two orders of magnitude.  To avoid such artifacts, we report all-electron results for $\omega$B97M(2) whenever the original FC recipe is demonstrably inadequate (e.g. HEAVY28\_13, HEAVY28\_14, CHB6\_3, CHB6\_6, MB16-43, AE11, MX34, and all TM-related sets).  

revDSD-PBEP86-D4 was originally developed using a distinct frozen-core (FC) protocol that is difficult to automate within our current workflow. Therefore, no FC approximation is applied for revDSD-PBEP86-D4 in this study, except for the AE11 and AE18 data sets. This treatment illustrates the potential pitfalls of a non-black-box approach that requires careful handling of the FC approximation. For reference, the WTMAD2 of revDSD-PBEP86-D4 is only about 25\% larger than that of $\omega$B97M(2) for the thermochemistry category in GMTKN55, whereas its mean NER on TC in the present work is roughly twice as large as $\omega$B97M(2). The primary source of this discrepancy is spin-symmetry breaking (discussed below), although the FC protocol also contributes significantly. Moreover, since the MP2 component in modern double-hybrid functionals—particularly RI-MP2 or localized RI-MP2—is often faster than the SCF stage, the FC approximation may not be essential for practical speedups and might best be avoided altogether in both training and production calculations. (We note, however, that this would require basis sets capable of describing core-valence and core correlation effects.) 

The most critical challenge for double-hybrid density functionals (DHDFTs) is spin contamination (SC), or in other words, spin-symmetry breaking (SSB), as has already been observed in some other studies~\cite{goerigk2011efficient,hait2018communication,liu2025revisiting}. A clear example of this issue arises for the 3B-69 data set, where revDSD-PBEP86-D4 shows abnormally large errors. The problem originates from the p-benzoquinone trimer: while the monomer is free of SC, both the dimer and trimer exhibit significant contamination, resulting in an error of about 20 kcal/mol, compared to less than 0.1 kcal/mol for PBE0-D4. To quantify its impact, we divided the 8,377 reactions into three groups according to the presence or absence of SC in r2SCAN0-D4 and revDSD-PBEP86-D4, and compared the “win rate” of revDSD-PBEP86-D4 relative to PBE0-D4 (i.e., the fraction of reactions where it yields a smaller error). For the 5,447 reactions free of SC in both functionals, the win rate is an impressive 78\%. This rate drops noticeably to 64\% for reactions where both functionals exhibit SC. The most severe degradation occurs for reactions in which only revDSD-PBEP86-D4 suffers from SC, likely caused by artificial symmetry breaking due to its high fraction of Hartree–Fock exchange (HFX); here, the win rate falls to just 38\%, indicating that this double hybrid can underperform its hybrid counterpart in such cases.

This trend is consistent with dataset-level behavior: revDSD-PBEP86-D4 clearly outperforms PBE0-D4 on single-reference sets such as TAE\_W4-17nonMR and BH28 but deteriorates significantly on strongly multireference systems, including TAE\_W4-17MR and ORBH36. Its poor performance on electric-field response properties also stems from this issue. In contrast, $\omega$B97M(2) is less susceptible because it uses $\omega$B97M-V orbitals, suggesting that xDH-type functionals whose orbitals are generated using a lower HFX content can substantially mitigate artificial SSB. Nonetheless, the improvement of $\omega$B97M(2) over $\omega$B97M-V on multireference-dominated data remains smaller than for single-reference cases. Emerging approaches such as the random phase approximation (RPA)~\cite{chen2017random} or regularized MP2 methods~\cite{shee2021regularized,carter2023repartitioned} behave better than MP2 in the presence of strong correlation, but cannot correct SSB in the orbitals. There is incentive to revisit the idea of orbital optimization (OO) in the presence of wavefunction correlation (the OO-DH approach\cite{peverati2013orbital}) or to include singles as well to restore broken orbital symmetries.

Our results illustrate both the current utility and present-day limitations of DH functionals. To ensure size-consistent results (because use of restricted orbitals is catastrophic for multireference problems or in heterolytic dissociations), one must use the unrestricted orbitals whenever they are lower in energy than restricted orbitals at the SCF level. This is what we have done here, with the benefit of guaranteed size-consistency, but with the drawback of reduced DH accuracy when either artificial or essential SSB occurs. Alternatively, in the regime of artificial SSB,\cite{lee2019distinguishing,liu2025revisiting} much better DH results \textit{could} be obtained by using the (unstable) restricted orbitals.  This is a user's choice that comes at the expense of the DH no longer obeying model chemistry precepts \cite{pople1999nobel}: potential energy curves may not be continuous, and results may not be size-consistent. While this is disturbing, it is just like the use of standard MP2 or even RPA or CCD, whose HF orbitals are highly susceptible to artificial SSB.\cite{lee2019distinguishing}  This is a manifestation of the famous symmetry dilemma of quantum chemistry.\cite{lowdin1963discussion} To minimize this problem, xDH methods with orbitals that do not suffer from extensive artificial SSB (such as $\omega$B97M(2)) are preferable, as this is the primary origin of its performance advantage over revDSD-PBEP86-D4. 
\section{Conclusions}\label{chap5sec:conclusion}

We have presented GSCDB137, a rigorously curated suite of 137 benchmark sets that roughly  doubles the data-point count of GMTKN55/MGCDB84 (to 8377 energies) while broadening the chemical and physical scope. New components include realistic transition-metal thermochemistry and kinetics, three-body non-covalent interactions, dipoles, polarizabilities, oriented-field response energies, and dimer vibrational frequencies. At the same time, legacy data have been re-evaluated, obsolete reference values replaced, and spin-contaminated or redundant entries removed. By gathering all of these elements under a single, fully documented umbrella, GSCDB137 furnishes a coherent, high-accuracy platform for functional assessment today and a robust training ground for the next generation of non-empirical or machine-learned density functionals.

Benchmarking 29 broadly used density functionals on this database confirms the expected Jacob’s-ladder hierarchy but also highlights important exceptions. Double hybrids remain the most accurate rung, with $\omega$B97M(2) and revDSD-PBEP86-D4 reducing the average error by roughly 30\% relative to their hybrid analogues. Among hybrids, $\omega$B97M-V emerges as the best HMGGA overall, while $\omega$B97X-V is the most balanced HGGA, especially for electric-field and vibrational properties. B3LYP only performs well for frequency calculations when used with at least a triple-zeta basis set; the commonly employed B3LYP/6-31G* combination remains inadequate, as shown in previous studies.\cite{liang2023analytical} B97M-V is the clear leader among meta-GGAs, and revPBE-D4 is the safest choice within the conventional GGA class. However, some categories violate the ladder ordering: semi-local r2SCAN-D4 outperforms many hybrids on frequencies, and EF errors show little correlation with other ground-state energetics. These findings warn against relying on energy-only training sets and underscore the need to include density-sensitive and response quantities in future functional design. From an applications standpoint, the present study suggests the following guidelines. For main-group chemistry or organometallic reactions, $\omega$B97M-V (or its D4 variant, or the runner-up, CF22D, where VV10 is unavailable) offers the best accuracy. CF22D is also a valuable choice for metal clusters. For frequencies, B3LYP-D4, PBE0-D4, and r2SCAN-D4 are recommended. 

Double-hybrid accuracy is impressive but not turnkey. We show that frozen-core protocols, basis-set completeness, and especially spin symmetry breaking character can each change double-hybrid errors by a large ratio (even orders of magnitude).  In practical calculations we therefore recommend (i) reproducing the training frozen-core definition whenever possible, (ii) using at least the training basis—def2-QZVPPD for $\omega$B97M(2)—and (iii) using a good hybrid functional to generate orbitals to avoid artificial symmetry breaking. When such precautions are unfeasible, a robust hybrid (e.g., $\omega$B97X-V or $\omega$B97M-V) may be the safer alternative at present.

\section*{Supporting Information}

The database files (calculation inputs and analysis scripts) are available on GitHub at: \\
\url{https://github.com/JiashuLiang/GSCDB}

Additional supporting files include:
\begin{itemize}
    \item \textbf{Raw Data:} \texttt{funcs.xlsx}
    \item \textbf{Functionals’ Error Statistics by Set:} \texttt{Errors\_per\_set.xlsx}
    \item \textbf{Relative Standard Error Ratios:} \texttt{Rel\_wrt\_hyb.xlsx}
\end{itemize}

\begin{acknowledgement}
This work was supported by the Director, Office of Science, Office of Basic Energy Sciences, of the U.S. Department of Energy through the Gas Phase Chemical Physics Program, under Contract No. DE-AC02-05CH11231. We thank the authors of MGCDB84, GMTKN55, and ACCDB for providing a solid foundation for this work. We are deeply grateful to Diptarka Hait for insightful discussions and valuable suggestions on refining the dipole moment and static polarizability data sets. We also thank Bun Chan for his generous guidance on the use of his benchmark sets and for providing so much highly accurate reference data. Our appreciation extends to Tarek Scheele and Tim Neudecker for clarifying details of the OEEF set. We also thank Klaas Giesbertz and Sebastian Ehlert for their suggestion on our preprint version of the database.

Most importantly, we thank all researchers who devote their time and effort to generating high-quality reference values, whether or not their work is included in this database. We fully recognize not only the substantial computational resources to complete high-level wavefunction calculations, as well as the human time required in the meticulous and often tedious process of evaluating the importance of each energy component and verifying the reliability of every data point. Such work can feel exhausting and thankless, lacking the allure of novelty, yet it is undeniably essential for progress. Without these efforts, we could neither rigorously assess the capabilities of current methods nor confidently guide the development of new ones. The creation of GSCDB137 thus also reflects the collective effort of a global community devoted to accuracy and reliability in computational chemistry; without their contributions, this database would simply not have been possible.
\end{acknowledgement}

\textbf{Competing interests: } M.H-G. is a part owner of Q-Chem Inc., whose software was used in the calculations reported here. 

\clearpage

\bibliography{reference}
\end{document}